# RESEARCH

# Comparative studies of the sensitivities of sparse and full geometries of Total-Body PET scanners built from crystals and plastic scintillators

M. Dadgar[1,2*], S. Parzych[1,2], J. Baran[1,2], N. Chug[1,2], C. Curceanu[4], E. Czerwiński[1,2], K. Dulski[1,2], K. Elyan[1,2], A. Gajos[1,2], B.C. Hiesmayr[5], Ł. Kapłon[1,2], K. Klimaszewski[6], P. Konieczka[6], G. Korcyl[1,2], T. Kozik[1], W. Krzemień[7], D. Kumar[1,2], S. Niedźwiecki, D. Panek[1,2], E. Perez del Rio[1,2], L. Raczyński[6], S. Sharma[1,2], Shivani[1,2], R.Y. Shopa[6], M. Skurzok[1,2], E. L. Stępień[1,2,3], F. Tayefi Ardebili[1,2], K. Tayefi Ardebili[1,2], S. Vandenberghe[8], W. Wiślicki,[6] and P. Moskal[1,2,3*]

*Correspondence:
meysam.dadgar@uj.edu.pl;
p.moskal@uj.edu.pl
1Faculty of Physics, Astronomy, and Applied Computer Science, Poland. Department of Experimental Particle Physics and Applications, Jagiellonian University, Krakow, Poland
2Total-Body Jagiellonian-PET Laboratory, Jagiellonian University, Krakow, Poland
3Theranostics Center, Jagiellonian University, Krakow, Poland
Full list of author information is available at the end of the article

**Abstract**

**Background:** Alongside the benefits of Total-Body imaging modalities, such as higher sensitivity, single-bed position, low dose imaging, etc., their final construction cost prevents worldwide utilization. The main aim of this study is to present a simulation-based comparison of the sensitivities of existing and currently developed tomographs to introduce a cost-efficient solution for constructing a Total-Body PET scanner based on plastic scintillators.

**Methods:** For the case of this study, eight tomographs based on the uEXPLORER configuration with different scintillator materials (BGO, LYSO), axial field-of-view (97.4 cm and 194.8 cm), and detector configuration (full and sparse) were simulated. In addition, 8 J-PET scanners with different configurations, such as various axial field-of-view (200 cm and 250 cm), the different cross-sections of plastic scintillator, and the multiple numbers of the plastic scintillator layers (2, 3, and 4), based on J-PET technology have been simulated by GATE software. Furthermore, Biograph Vision has been simulated to compare the results with standard PET scans. Two types of simulations have been performed. The first one with a centrally located source with a diameter of 1mm and a length of 250 cm, and the second one with the same source inside a water-filled cylindrical phantom with a diameter of 20 cm and a length of 183 cm.

**Results:** With regards to sensitivity, among all the proposed scanners, the ones constructed with BGO crystals give the best performance ($\sim$ 350 cps/kBq at the center). The utilization of sparse geometry or LYSO crystals significantly lowers the achievable sensitivity of such systems. The J-PET design gives a similar sensitivity to the sparse LYSO crystal-based detectors while having full detector coverage over the body. Moreover, it provides uniform sensitivity over the body with additional gain on its sides and provides the possibility for high-quality brain imaging.

**Conclusion:** Taking into account not only the sensitivity but also the price of the Total-Body PET tomographs, which till now was one of the main obstacles in their widespread clinical availability, the J-PET tomography system based on plastic scintillators could be a cost-efficient alternative for Total-Body PET scanners.

**Background**

Positron Emission Tomography (PET) is the most advanced and sensitive nuclear medicine imaging modality established for oncology applications [1]. Total-Body (TB) PET scanners extended the applicability of molecular imaging to a wider range of fields such as cardiovascular disease, multi-organ imaging, physiological study, treatment monitoring, whole-body dynamic imaging, etc., which are not possible with current tomographs due to their limited axial field-of-view (AFOV) [2–6]. TB PET scanners can deliver excellent new diagnostics features such as parametric imaging and dynamic imaging resulting in better specificity for distinguishing between cancer and inflammations [7–9].

The first TB PET, the uExplorer system, with 194.8 cm of AFOV, has been constructed at UC Davis in California using crystal scintillators [10–12]. uExplorer has been constructed based on the utilized technology in conventional PET tomographs and achieves its large AFOV by equipping the scanner with multiple units with large amounts of scintillation crystals, SiPM, and electronics, which significantly increases the construction price compared to current clinical PET scanners as shown in Figure 1.

The utilized detection principle based on the current clinical tomographs and the radial arrangement of scintillation crystals and SiPMs in uEXPLORER causes a higher price per unit compared to the conventional PET. As a consequence, high construction cost is the main obstacle preventing uEXPLORER from broad dissemination in hospitals [7, 13, 14].

Alongside the wide range of applications and benefits of TB PET imaging, the high cost of employing per unit reduces the hopes for comprehensive usage in clinics. This problem emphasizes the demand for alternative technology for developing affordable TB PET to make it publicly available in clinics [7, 15].

To map the substantial consequences of the high construction cost of TB PET with traditional technology, it is essential to consider it as the main barrier to the possible function of these types of scanners in medical research clinics. A cost-efficient alternative solution for TB PET construction can open new horizons for rapid research development, such as drug delivery, radiopharmaceutical investigations, treatment planning, etc [16, 17].

All the mentioned facts motivate the investigation to find a cost-effective solution to reduce the final construction cost of TB PET [7]. One of the recently suggested solutions is the sparse detector configuration [18,19]. In the proposed sparse geometry, there are intervals between detectors leading to large AFOV with less number of scintillation crystals and electronics. However, sparse geometry is unable to provide uniform sensitivity in all parts of the patient's body due to the gaps, as shown in Figure 1.

Utilization of BGO scintillation crystals instead of LYSO in the presented uEXPLORER TB PET is the other recommended solution since BGO is 2 to 3 times less costly than LYSO. Using BGO reduces the crystal part of the costs, while the total number of SiPMs and electronics are similar in both cases, which will not make a major reduction in the construction price of the TB PET system [7, 20]. Jagiellonian Positron Emission Tomography (J-PET) is an advanced scanner technology developed over a decade and constructed from plastic scintillators. One of the main differences between J-PET and other traditional tomography is an axial arrangement of detection panels, as shown in Figure 2, while the common PET is constructed with radial arrangements of detectors [21–26].

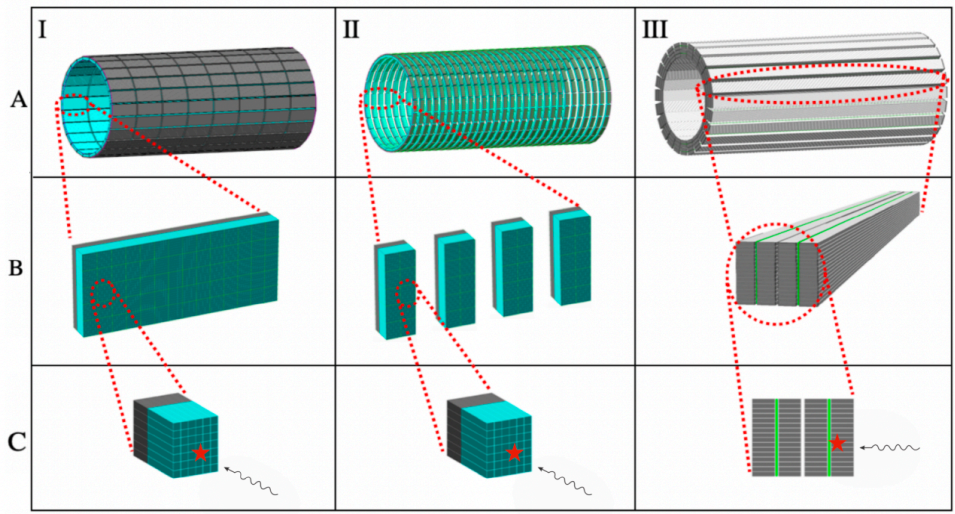

**Figure 1** A.I, Visualization of uEXPLORER TB PET scanner consisting of 8 detection units (rings). Each unit is made up of 24 detection blocks (B.I). The blocks in uEXPLORER are composed of arrays (C.I). The arrays are the smallest detection units of uEXPLORER and hold 6 x 7 scintillation crystals (blue) together and couple them to the SiPM matrix (black). A.II shows the sparse geometry based on uEXPLORER, including 29 detection units (rings). In contrast, arrays are distributed in smaller detection blocks (B.II). A.III TB J-PET consists of 4 layers of EJ-230 plastic scintillator strips (grey), which have been located parallel to the axial axes of the scanner (B.III). The modules in TB J-PET have been equipped with wavelength-shifting plastics (WLS, green).

Due to the unique geometrical configuration applied in J-PET technology, it employs different detection principles of gamma quanta, and localization methods of annihilation points [21, 27–29]. The interaction position of the gamma quanta with plastic scintillators is determined by the arrival time of the light signals to each end of the scintillation strips (Fig. 2 left).

Considering the distinct arrangement of the detection panels, J-PET technology enables a cost-efficient PET for TB PET imaging with up to 2.5 m long AFOV. The main aim of the presented simulation-based study is to compare the sensitivity of existing and presently developed TB PET imaging modalities.

**Methods**

The study presented in this article was carried out using GATE (Geant4 Application for Tomographic Emission) software. GATE is a validated simulation tool based on the Monte-Carlo method dedicated to nuclear medicine applications [30, 31]. The uEXPLORER and the J-PET presented their solution to achieve TB PET scans. Alongside similar applications of these technologies, they use different materials and designs in their scanner. The two main parameters that distinguish these technologies are the scintillator material (organic plastics or inorganic crystals) and the arrangement of detection units. For each of the groups, several tomographs have been simulated to be compared based on all parameters which influence their sensitivity and cost.

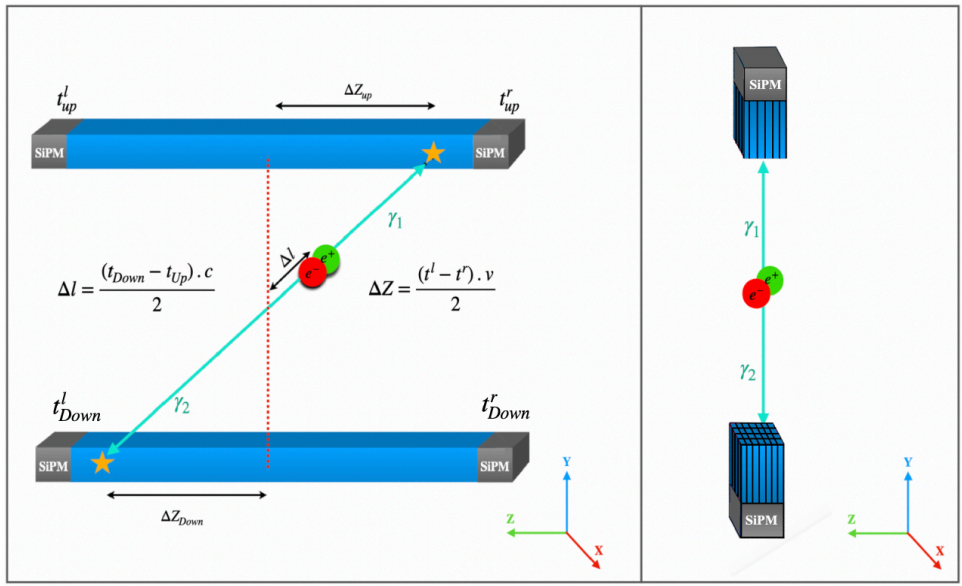

**Figure 2** Illustration of scintillators arrangements, the principle of detection and reconstruction of annihilation positions. (Left) Representation of J-PET plastic scintillator-based technology where the axially arranged scintillator (blue) is readout by two photomultipliers of both ends (gray). (Right) Representation of crystal-based technology with the radially arranged scintillators (blue), where each one is read out with corresponding photomultipliers (gray). The fractional energy resolution for the energy deposited by the annihilation photon in a single plastic scintillator strip has been measured $\sigma(E)/E \simeq 0.044/\sqrt{E(MeV)}$, based on the previous investigation in J-PET

The uEXPLORER

The uEXPLORER-based scanners are the first group of tomographs that were simulated. For these cases, we simulated 8 PET scanners based on the crystal configuration of the uEXPLORER. These groups stand for current clinical tomographs that have a radial arrangement of scintillation crystals. These eight configurations, combined from 2 different types of crystals (LYSO and BGO), two different AFOV (194.8 and 97.4 cm), and two different geometrical configurations (full and sparse) were simulated [8, 9, 32]. The reasons for performing an investigation over the scanners mentioned above are the proposed solutions to reduce the construction cost of TB PET scanners. Sparse geometry has been introduced as a cost-effective solution to extend AFOV [33].

The J-PET

In this study, we simulated 8 TB J-PET scanners with various configurations of panels to evaluate the effect of plastic scintillator dimensions and multiple layers of modules on tomograph performance. For the scintillator with a cross-section of 4 mm x 20 mm, a two-layer geometry with 200 cm and 250 cm lengths was simulated. For the plastic strips with a 6 mm x 30 mm cross-section, two, three, and four layers configurations with 200 cm and 250 cm lengths were simulated. Plastic scintillator strips have been equipped with SiPM at each end in all panel configurations. These specific arrangements of SiPMs and plastic scintillators allow for extending the AFOV without incrementing the number of SiPMs or electronics but only by increasing the length of plastic strips [7].

The detection panels of TB J-PET are equipped with wavelength shifters (WLS) which are utilized to improve axial resolution [20,26,34]. The WLS with dimensions of 3 mm x 108.15 mm x 6 mm is located perpendicular to the plastic strips, as shown in Figure 1. The WLS layers are read out by the SiPMs, coupled to them from one side.

In the TB J-PET, the expected axial spatial resolution for the registration of gamma photons is equal to sigma = 2.1 mm. Spatial resolutions of the image of 3.7mm (transversal) and 4.9mm (axial) are estimated [20]. The TOF resolution for the J-PET was estimated as a function of the lengths of the applied scintillator strips, and it varies from CRT=140ps to CRT=240ps when the length is increased from 50 cm to 200 cm [20]. For the scatter fraction reduction, the energy loss threshold of 200 keV will be used, resulting in the scatter fraction of 36.2 % [20]. For the two-layer total-body J-PET solution, the noise equivalent count rate NECR peak was estimated to be 630 kcps at kBq/mL [20], which is in between the values obtained by uEXPLORER (1524 kcps at 17.3 kBq/mL) [8] and Biograph Vision (306 kcps at 32 kBq/mL) [35].

Biograph Vision scanners

In order to compare the results of TB PET scanners to standard ones, the Biograph Vision from Siemens was simulated [35–37]. Biograph Vision is composed of 8 rings, where each ring consists of 38 panels. Each panel is built from a 20 x 10 array of 3.2 x 3.2 x 20 mm LSO crystals, providing 32 mm in axial direction [35]. In total, Biograph Vision spans 26.3 cm AFOV.

For each one of the geometries, as mentioned earlier, two types of simulations have been performed, (i) with a line source of the diameter of 1mm and a length of 250 cm located along the central axis of the tomograph and (ii) with a line source surrounded by a cylindrical water-filled phantom with a diameter of 20 cm and a length of 183 cm.

The sensitivity for a slice ($S_i$) was calculated according to the following formula:

$$S_i = \frac{R_i \times L_{mean}}{d \times A_{mean}} \quad , \tag{1}$$

Where $L_{mean}$ is the source length, $d$ is the width of the slice, and $A_{mean}$ is the initial activity. The rate $R_i$ of each slice in counts per second is determined by dividing the counts collected in the slice by the duration of the measurement.

TB PET provides extended AFOV that considerably improves sensitivity compared to the current clinical PET. Still, it is required to determine new event selection criteria to achieve optimum results [13].

TB PET scans, thanks to the larger AFOV, are capable of detecting more oblique coincidences. While these events contribute positively to the increase of system sensitivity, they deteriorate the axial resolution of the tomograph [38]. Since sensitivity and spatial resolutions are the main characteristics of PET, making a trade-off between these two parameters will enhance the quality of the final reconstructed image [39]. The optimization is performed as a function of the acceptance angle (shown in Figure 3), which is used for pre-selecting those maximum azimuthal coincidences contributing to the reconstructed image.

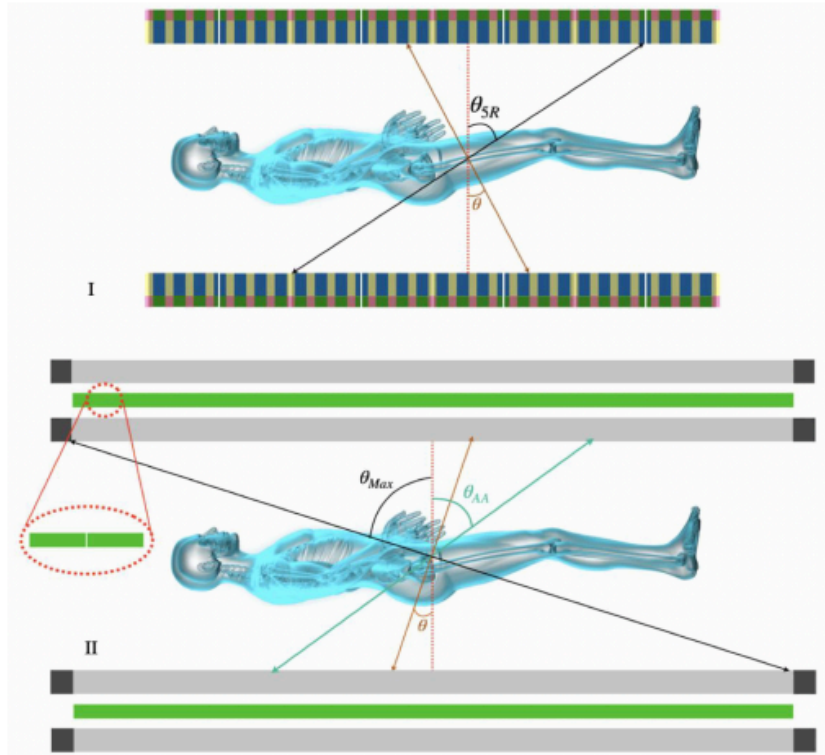

**Figure 3** Schematic visualization of uEXPLORER (I) (blue) [8, 10, 13]) and sparse geometry (transparent yellow) with 194.8 cm of AFOV and dual layers. TB J-PET (II) with axially arranged plastic scintillators (grey) coupled with SiPMs (black) at each end and arrays of WLS (green) between each layer. The oblique LORs (with large values of θ) have a negative contribution in the spatial resolutions due to the parallax error. To avoid it, uEXPLORER uses a ring-based cut that accepts the coincidences within a maximum of 5 rings. As shown in figure (II), TB J-PET uses continuous plastic scintillators (grey), $\theta_{AA}$ denotes the acceptance angle applied for it to cut oblique LORs. $\theta_{Max}$ demonstrate the largest angle of detectable oblique coincidences.

Due to the essential difference between the geometrical configurations of TB JPET and other TB PET scanners, it is necessary to define distinct logic for acceptance angle cuts for them. The TB J-PET provides 2.5 m AFOV with one detection ring. At the same time, the rest of the configurations described in the " The uEXPLORER" sub-section are constructed with several rings along the axial axis. Acceptance angle cut ($\theta_{AA}$) is a suitable choice for the case of TB J-PET (Figure 3), but for other configurations, the results will be presented based on maximum ring difference [40].

In this study, two sets of simulations have been performed, first only with a line source axially located at the center of the scanner, then with the same source while surrounded by a cylindrical water-filled phantom. Despite many advantages in performing TB imaging, applying cuts to suppress the oblique coincidence detection is essential to avoid their destructive effect on the spatial resolution of tomographs. Accordingly, the sequences of such cuts on the sensitivity of dedicated scanners have been investigated. For all the scanners, 57 degrees of acceptance angle cut or its equivalent has been performed. For scanners that utilize crystal scintillators, the equivalent is represented by a ring-wise cut.

The acceptance cut has a significant effect on the reduction of the sensitivity of TB PET scanners. Tomographs such as Biograph Vision and 97.4 cm uEXPLORERs' were neutral against this cut, which can be explained due to their smaller AFOV, which fits inside the cut region. Moreover, the effect of this cut has been investigated in the presence of the phantom to mimic clinical scenarios.

The Total-Body sensitivity is defined as the average of rate of detected annihilations originating from within the 183 cm long phantom (which mimics a human body) divided by the total activity $A_{body}$ present within it:

$$S_{TB} = \frac{\Sigma_{i=1}^{N} R_i}{A_{body} \times N} \quad , \tag{2}$$

Where N is the number of slices within the body range, in the case of scanners with an axial field-of-view shorter than 183 cm, the empty slices are outside of the scanner but still within the body range and are also taken into averaging.

**Results**

The sensitivity profiles of the scanners described in the method section based on a 2.5 m line source located in their central axial axes have been shown in Figure 4. It includes all the sensitivity profiles of the scanners which have been investigated in this study.

As shown in Figure 4, the full uEXPLORER (BGO) with 194.8cm of AFOV provides higher sensitivity compared to the other scanner. Among the J-PET investigated scanners, TB J-PET (4 modules), based on the plastic scintillator with a cross-section of 6 × 30 mm and 200 cm of AFOV, has higher sensitivity. Among sensitivity profiles presented in Figure 4 the TB PET scanners based on a sparse configuration have a wavy sensitivity profile, which is due to their geometrical configuration that has not full coverage of the detector along their AFOV.

As described in the method section, the utilization of a cut to suppress the oblique coincidences is required in TB PET scanners. To show the effect of this cut on the sensitivity profiles of the investigated TB PET, a 57 degree of acceptance angle cut and a maximum five ring difference cut has been applied for J-PET-based scanners and uEXPLORER tomographs, respectively. These cuts have been applied based on the previously investigated study by these research groups [40].

As shown in Figure 5, applying these cuts reduces the sensitivity of the scanner at the regions close to the center of the scanner. However acceptance angle cut applied in the J-PET-based scanner provides a uniform region in the corresponding sensitivity profile, while the maximum ring difference cut applied in the uEXPLORER-based scanner causing a wavy region in their sensitivity profiles. As shown in Figure 5, applying five ring difference cut does not have an influence on the scanner, such as Full uEXPLORER BGO/ LYSO with an AFOV of 97.4 cm. The reason for such independent behavior of those scanners is due to the smaller AFOV, which is not influenced by five ring difference cut.

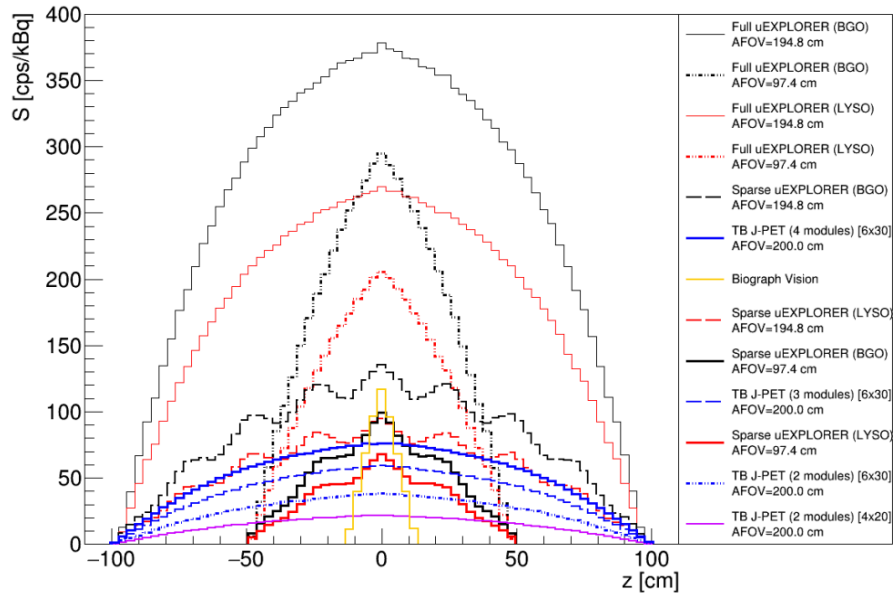

**Figure 4** Sensitivity [cps/kBq] profiles of all the TB PET geometries with a 2.5 m line source located in their central axial axis.

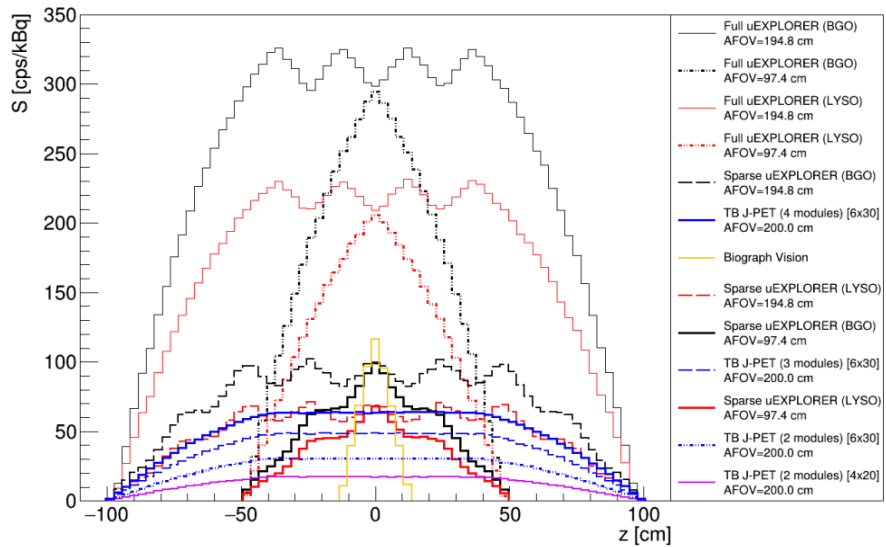

**Figure 5** Sensitivity [cps/kBq] profiles of all the TB PET geometries with a 2.5 m line source located in their central axial axis.

Figure 6 shows the sensitivity profiles of the TB PET scanners with a 2.5 m centrally located line source inside a 183 cm water-filled phantom with a diameter of 20 cm. Among all TB PET scanners, the uEXPLORER based on the BGO with 194.8 cm of AFOV has higher sensitivity, and among J-PET based scanners, TB J-PET (4 modules), based on the plastic scintillator with a cross-section of 6 × 30 mm and 200 cm of AFOV provides higher sensitivity.

The effect of applying a cut to suppress oblique coincidences has been shown in Figure 7. Although, in the uEXPLORER-based TB PET, the sensitivity profiles after applying five ring difference cut have less fluctuated in the presence of phantom, the J-PET-based scanners provide a more uniform region even after applying 57 degrees of acceptance angle.

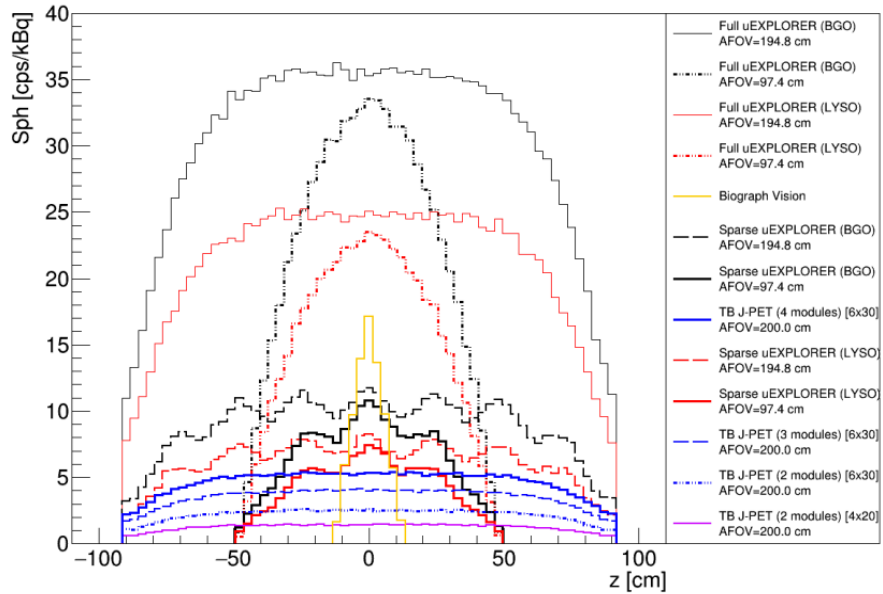

**Figure 6** Sensitivity [cps/kBq] profiles of the TB PET scanners with a 2.5 m line source inside a 183 cm water-filled cylindrical phantom with a diameter of 20cm.

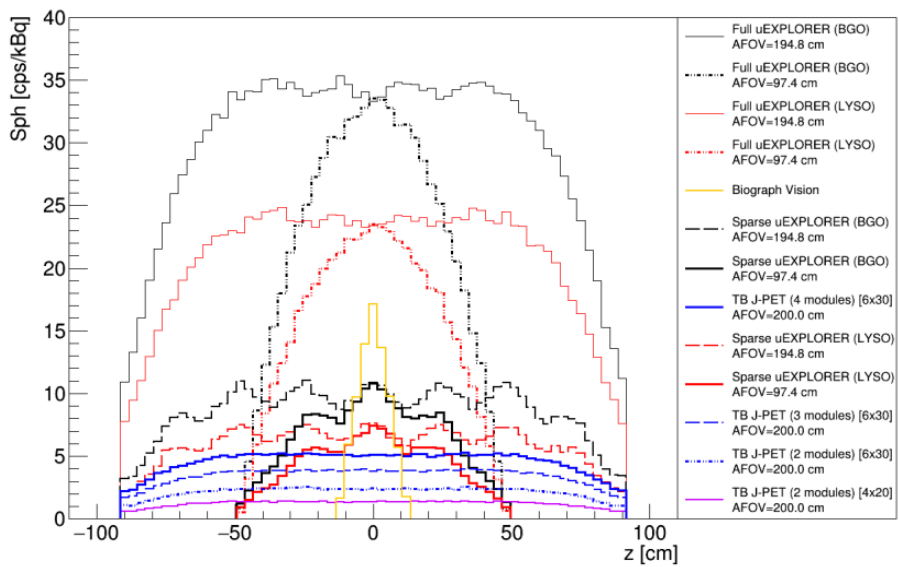

**Figure 7** Sensitivity [[cps/kBq] profiles of the TB PET scanners with a 2.5 m line source inside a 183 cm water-filled cylindrical phantom with a diameter of 20 cm, with 57o degree of acceptance angle for J-PET-based scanners and five ring difference cut for uEXPLORER tomographs

**Discussion**

The main aim of this work was to perform a simulation study to compare several cost-efficient solutions with TB J-PET scanners based on sensitivity, which is one of the main characteristics of tomographs. Generally, the cost of manufacturing a TB PET scanner depends on two main factors: the price of the scintillators and the readout electronics, including SiPMs. With construction costs estimated to be at the level of 10 million USD [13], the 194.8 cm uEXPLORER will be used as a reference in the following considerations. Because a significant portion of the system price is constituted by the scintillators, using BGO instead of LYSO can reduce this share of the total cost by 30%.

However, due to the same number of SiPMs and electronic readouts, the cost of making a tomograph will remain relatively high. Construction of TB PET by utilization of sparse configuration of the detector has been proposed as another solution investigated in this work. As shown in Figure 9, the sensitivity of these scanners is almost one-third of full uEXPLORER tomographs. However, they have nearly half a photomultiplier covered area. As shown in Figure 3 and 9, the technology used in J-PET makes it possible to build scanners with higher AFOVs with only a slight increase in the number of SiPMs, which are due to the addition of WLS readouts [34]. With this assumption, we simulated TB J-PET designs with AFOV reaching as high as 250 cm. Figure 8 shows the sensitivity profiles of the alternative TB PET scanners compared to proposed J-PET-based systems, with a cylindrical phantom with a diameter of 20cm and a length of 183cm.

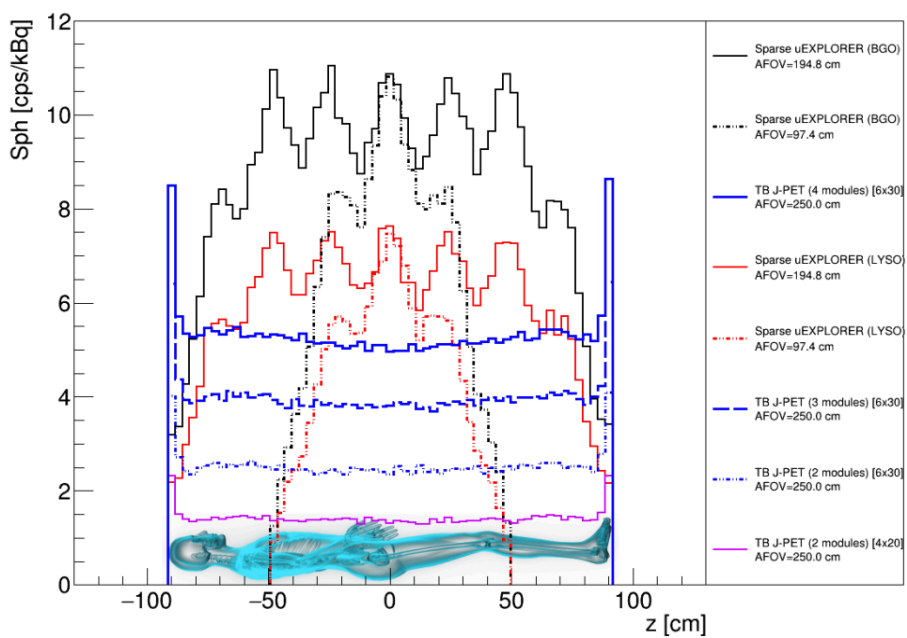

**Figure 8** Sensitivity [cps/kBq] profile of the TB PET scanners with a 2.5 m line source inside a 183 cm long water-filled cylindrical phantom with a diameter of 20 cm and acceptance cut. The schematic visualization of a male patient is added to represent the position of a 183 cm height person in the sensitivity profile of the tomographs with 57 degree of acceptance angle for J-PET-based scanners and five ring difference cut for uEXPLORER tomographs.

For better visualization, a schematic image of a 183 cm tall male person was located in the position of the phantom. The sensitivity profiles of all J-PET proposed TB PET scanners show even higher sensitivity in the brain region, enhancing their performance for simultaneous brain and body scans. This is caused by a reduced attenuation material for one of the annihilation photons while still having more than 30 cm of the sensitive part of the tomograph. This shows that TB J-PET scanners with 250 cm of AFOV can provide higher sensitivity in positions of vital organs while using plastic scintillators, which are less expensive than crystal scintillators.

As shown in Figure 9, the TB PET based on J-PET technology has a smaller photomultiplier covered area while having comparable total sensitivity to other tomographs. According to a study performed by Moskal et al. [20], which investigated the utilization of 50 cm long plastic strips, such dimensions can provide a good compromise between spatial characteristics and the construction price of TB J-PET.

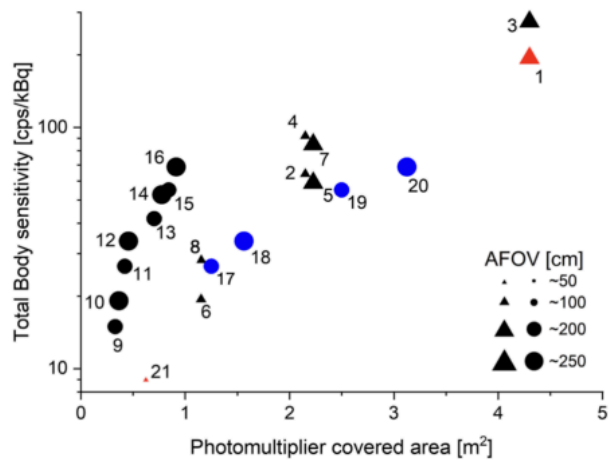

| Nr | Scanner | AFOV [cm] | TB-FOM | | | | |
|----|---------|-----------|--------|----|-----------------------------------|-------|------|
| 1  | uEXPLORER (LYSO) | 194.8 | 1 | 11 | TB J-PET (2 modules) [6x30 mm] | 200 | 2.74 |
| 2  |         | 97.4  | 0.66 | 12 |                                   | 250 | 3.18 |
| 3  | uEXPLORER (BGO) | 194.8 | 2.13 | 13 | TB J-PET (3 modules) [6x30 mm] | 200 | 2.56 |
| 4  |         | 97.4  | 1.44 | 14 |                                   | 250 | 2.88 |
| 5  | Sparse uEXPLORER (LYSO) | 194.8 | 0.58 | 15 | TB J-PET (4 modules) [6x30 mm] | 200 | 2.74 |
| 6  |         | 97.4  | 0.37 | 16 |                                   | 250 | 3.15 |
| 7  | Sparse uEXPLORER (BGO) | 194.8 | 1.27 | 17 | TB J-PET (2 modules & 4/5 rings) [6x30 mm] | 4x50 | 0.93 |
| 8  |         | 97.4  | 0.80 | 18 |                                   | 5x50 | 0.96 |
| 9  | TB J-PET (2 modules) [4x20 mm] | 200 | 1.94 | 19 | TB J-PET (4 modules & 4/5 rings) [6x30 mm] | 4x50 | 0.96 |
| 10 |         | 250 | 2.18 | 20 |                                   | 5x50 | 0.96 |
|    |         |     |      | 21 | Biograph Vision                   | 26.3 | 0.62 |

**Figure 9** Representation of the total sensitivity of the scanner as a function of the sum of the required photomultiplier covered area. The TB PET systems based on the J-PET technology are marked with black circles. Four additional ring-based TB J-PET geometries are marked with blue circles. The crystal-based geometries are marked in the plot with black triangles. Each ring of the J-PET is constructed as a separate tomograph with 50 cm length scintillators. Two existing systems (Biograph Vision and uEXPLORER from LYSO crystals) were colored red..

Due to the fact that currently, existing organic scintillators have limitations of light transfer characteristics [26, 41], a TB J-PET design based on the ring-wise configuration is being explored. With that in mind, four additional TB J-PET geometries are presented in Figure 9. The ring-wise configuration does not influence the corresponding Total-Body sensitivity since the AFOV of the scanners remains the same.

For the comparison of various discussed solutions, taking into account the sensitivity and costs of the systems, we introduce a figure of merit for the whole-body PET as:

$$TB - FOM = \frac{S_{TB}}{costs} \quad (3)$$

Where $S_{TB}$ denotes the average sensitivity of the system for the whole-body imaging, as defined in eq. 2. The relative costs were estimated taking into account that approximately half of the price of the PET system is due to the photomultipliers and electronics and another half due to the scintillators [2]. The result normalized to the Explorer is indicated in the table in Figure 9. It shows that the best value for total-body PET when considering both cost and sensitivity is the 2-layer PET based on plastic scintillators strips.

**Conclusions**

The remarkable results of uEXPLORER, along with studies presented in parallel by other research groups, have popularized the development of TB systems [1–3,10,42]. In this regard, the cost of manufacturing and deploying TB PET based on uEXPLORER limits the possibility for widespread use by diagnostic, therapeutic, and research centers. Nowadays, most TB PET developers accept these systems' high construction costs as a major barrier and are looking for cost-effective alternatives. The presented simulation-based study has been performed by pointing to the construction price of TB PET scanners as the most crucial obstacle in their development and widespread utilization. The sensitivity of scanners has been used as a metric to compare all possible solutions to construct cost-efficient TB PET scanners. The proposed tomographs based on J-PET technology by utilization of novel detector arrangement and unique detection principles (Figure 1) can be costefficient alternative TB PET scanners. As an example, the TB J-PET (4 modules and 250 cm of AFOV) needs about five times less amount of SiPM and electronic readouts (as shown in Figure 9). Based on this considerable reduction of the scintillator price and a smaller amount of SiPMs compared to the uEXPLORER, it can be claimed that J-PET technology has promising performance both from imaging characteristics [20] and being cost-efficient point of views.

**Acknowledgements**
Not applicable.

**Funding**
This work was supported by Foundation for Polish Science through TEAM POIR .04.04.00 – 00 – 4204/17, the National Science Centre, Poland (NCN) through grant No. 2021/42/A/ST2/00423, PRELUDIUM 19, agreement No. UMO – 2020/37/N/NZ7/04106 and the Ministry of Education and Science under the grant No. SPUB/SP/530054/2022. The publication also has been supported by a grant from the SciMat and qLife Priority Research Areas under the Strategic Programme Excellence Initiative at the Jagiellonian University.

## Abbreviations
J-PET: Jagiellonian Positron Emission Tomography; TB: Total-Body; PET: Positron Emission Tomography; AFOV: Axial field-of-view; LYSO: Lutetium–yttrium oxyorthosilicate; BGO: Bismuth germanium oxide; GATE: Geant4 Application for Tomographic Emission; SiPM: Silicon photomultipliers; PMT: Photomultiplier tubes.

## Availability of data and materials
The data used and/or analysed during the current study are available from the corresponding author upon request.

## Ethics approval and consent to participate
Not applicable.

## Competing interests
The authors declare that they have no competing interests.

## Consent for publication
Not applicable.

## Authors' contributions
All authors contributed to the development of methods for data analysis, interpretation of results, manuscript revision and approval. Simulation of utilized systems and analysis methods were mainly developed by MD and SP along with PM. MD, SP and PM edited the manuscript. All authors read and approved the final manuscript.

## Author details
[1]Faculty of Physics, Astronomy, and Applied Computer Science, Poland. Department of Experimental Particle Physics and Applications, Jagiellonian University, Krakow, Poland. [2]Total-Body Jagiellonian-PET Laboratory, Jagiellonian University, Krakow, Poland. [3]Theranostics Center, Jagiellonian University, Krakow, Poland. [4]INFN, Laboratori Nazionali di Frascati, Frascati, Italy. [5]Faculty of Physics, University of Vienna, Vienna, Austria. [6]Department of Complex Systems, National Centre for Nuclear Research, Otwock-Swierk, Poland. [7]High Energy Physics Division, National Centre for Nuclear Research, Otwock-Swierk, Poland. [8]Department of Electronics and Information Systems, MEDISIP, MEDISIP, Ghent University-IBiTech, Gent, Belgium.

## References
1. Alavi, A., Werner, J., Stepien, E., Moskal, P.: Unparalleled and revolutionary impact of pet imaging on research and day to day practice of medicine. Bio-Algorithms and Med-Systems 17(4), 203–212 (2021). doi:10.1515/bams-2021-0186.
2. Vandenberghe, S., Moskal, P., Karp, J.S.: State of the art in total body pet. EJNMMI Phys 7:35 (2020). doi.org/10.1186/s40658-020-00290-2.
3. Vandenberghe, S.: Progress and perspectives in total body pet systems instrumentation. Bio-Algorithms and Med-Systems 17(4), 265–267 (2021). doi:10.1515/bams-2021-0187.
4. Alavi, A., Saboury, B., Nardo, L.: Potential and most relevant applications of total body pet/ct imaging. Clinical Nuclear Medicine 47(1), 43–55 (2022). doi: 10.1097/RLU.0000000000003962
5. Surti, S., Pantel, P., Karp, J.: Total body pet: Why, how, what for? IEEE Transactions on Radiation and Plasma Medical Sciences 4(3), 283–292 (2020). doi:10.1109/TRPMS.2020.2985403.
6. Cesar, M., Todd, T., Phil, S.: Low-dose imaging in a new preclinical total-body pet/ct scanner. Frontiers in Medicine 6:88 (2019). doi:10.3389/fmed.2019.00088.
7. Moskal, P., Stepien, E.: Prospects and clinical perspectives of total-body pet imaging using plastic scintillators. PET Clin 15(4), 439–452 (2020). doi: 10.1016/j.cpet.2020.06.009.
8. Spencer, B., Berg, E., Schmall, J., et al: Performance evaluation of the uexplorer total-body pet/ct scanner based on nema nu 2-2018 with additional tests to characterize pet scanners with a long axial field of view. J Nucl Med 62(6), 861–870 (2021). doi:10.2967/jnumed.120.250597.
9. Leung, E., Berg, E., Omidvari, N., et al.: Quantitative accuracy in total-body imaging using the uexplorer pet/ct scanner. Phys Med Biol 66(20), 205008 (2021). doi:10.1088/1361-6560/ac287c.

10. Badawi, R.D., Shi, H., Hu, P., et al.: First human imaging studies with the explorer total-body pet scanner. J Nucl Med 60(3), 299–303 (2019). doi: 10.2967/jnumed.119.226498.
11. Lan, X., Fan, K., Li, K., Cai, W.: Dynamic pet imaging with ultra-low-activity of 18f-fdg: unleashing the potential of total-body pet. Eur J Nucl Med Mol Imaging 48(13), 4138–4141 (2021). doi: 10.1007/s00259-021-05214-5
12. Wang, Y., Li, E., Cherry, S., Wang, G.: Total-body pet kinetic modeling and potential opportunities using deep learning. PET Clin 16(4), 613–625 (2021). doi: 10.1016/j.cpet.2021.06.009.
13. Cherry, S., Jones, T., Karp, J., et al.: Total-body pet: Maximizing sensitivity to create new opportunities for clinical research and patient care. J Nucl Med 59(1), 3–12 (2018). doi:10.2967/jnumed.116.184028.
14. Poon, J., Dahlbom, M., Moses, W., et al.: Optimal whole-body pet scanner configurations for different volumes of lso scintillator: a simulation study. Phys Med Biol 57(13), 4077–4094 (2012). doi:10.1088/0031-9155/57/13/4077.
15. Efthimiou, N.: New challenges for pet image reconstruction for total-body imaging. PET Clin 15(4), 439–452 (2020). doi:10.1016/j.cpet.2020.06.002.
16. Moskal, P., Stepien, E.: New trends in theranostics. Bio-Algorithms and Med-Systems 17(4), 199–202 (2021). doi:10.1515/bams-2021-0204
17. Moskal, P., Rzaca, C., Stepien, E.: Novel biomarker and drug delivery systems for theranostics – extracellular vesicles. Bio-Algorithms and Med-Systems 17(4), 301–309 (2021). doi:10.1515/bams-2021-0183
18. Zein, S., Karakatsanis, N., Issa, M., et al.: Physical performance of a long axial field-of-view pet scanner prototype with sparse rings configuration: A monte carlo simulation study. Med Phys 47(4), 1949–1957 (2020). doi: 10.1002/mp.14046.
19. Karakatsanis, N., Nehmeh, M., Conti, M., et al.: Physical performance of adaptive axial fov pet scanners with a sparse detector block rings or a checkerboard configuration. Phys. Med. Biol. 67, 105010 (2022). doi:0.1088/1361-6560/ac6aa1
20. Moskal, P., Kowalski, P., Shopa, R., et al.: Simulating nema characteristics of the modular total-body j-pet scanner—an economic total-body pet from plastic scintillators. Phys Med Biol 66(17), 175015 (2021). doi:10.1088/1361-6560/ac16bd.
21. Moskal, P., Niedzwiecki, S., Bednarski, T., Czerwinski, E., Kaplon, , et al.: Test of a single module of the j-pet scanner based on plastic scintillators. Nucl. Instr. and Meth. A 764, 317–321 (2014). doi:10.1016/j.nima.2014.07.052
22. Moskal, P., Gajos, A., Mohammed, M., et al.: Testing cpt symmetry in ortho-positronium decays with positronium annihilation tomography. Nat Commun 12(5658) (2021). doi:10.1038/s41467-021-25905-9.
23. Moskal, P., Dulski, K., Chug, N., Curceanu, C., Czerwi´nski, E., et al.: Positronium imaging with the novel multiphoton pet scanner. Sci. Adv. 7(42), 4394 (2021). doi:10.1126/sciadv.abh4394.
24. Moskal, P., Kisielewska, D., Curceanu, C., et al.: Feasibility study of the positronium imaging with the j-pet tomograph. Phys Med Biol 7(64), 055017 (2019). doi: 10.1088/1361-6560/aafe20.
25. Kowalski, P., Wislicki, W., Shopa, R., et al.: Estimating the nema characteristics of the j-pet tomograph using the gate package. Phys Med Biol 63(165008) (2018). doi:10.1088/1361-6560/aad29b.
26. Moskal, P., Rundel, O., Alfs, D., et al.: Time resolution of the plastic scintillator strips with matrix photomultiplier readout for j-pet tomograph. Phys Med Biol 61(5), 2025–47 (2018). doi: 10.1088/0031-9155/61/5/2025.
27. Moskal, P., Zon, N., Bednarski, T., Bia las, P., Czerwinski, E., et al.: A novel method for the line-of-response and time-of-flight reconstruction in tof-pet detectors based on a library of synchronized model signals. Nucl. Instr. and Meth. A 775, 54–62 (2015). doi:10.1016/j.nima.2014.12.005
28. Raczynski, L., Moskal, P., Kowalski, P., Wislicki, W., Bednarski, T., Bia las, P., et al.: Novel method for hit-position reconstruction using voltage signals in plastic scintillators and its application to positron emission tomography. Nucl. Instr. and Meth. A 764, 186–192 (2015). doi:10.1016/j.nima.2014.07.032
29. Raczynski, L., Wislicki, W., Klimaszewski, K., Krzemien, W., Kopka, P., et al.: 3d tof-pet image reconstruction using total variation regularization. Physica Medica 80, 230–242 (2020). doi:10.1016/j.ejmp.2020.10.011
30. Jan, S., Santin, G., Strul, D., et al: Gate: a simulation toolkit for pet and spect. Phys Med Biol 49(19), 4543 (2004). doi: 10.1088/0031-9155/49/19/007.

31. Sarrut, D., Bala, M., Bardies, M., Bert, J., et al: Advanced monte carlo simulations of emission tomography imaging systems with gate. Phys Med Biol 66(10TR03) (2021). doi:10.1088/1361-6560/abf276.
32. Zhang, X., Xie, Z., Berg, E., et al: Total-body dynamic reconstruction and parametric imaging on the uexplorer. J Nucl Med 61(2), 285–291 (2021). doi:10.2967/jnumed.119.230565.
33. Zhang, J., Knopp, M., MV, K.: Sparse detector configuration in sipm digital photon countingvpet: a feasibility study. Mol Imaging Biol 21(3), 447–453 (2018). doi:10.1007/s11307-018-1250-7.
34. Smyrski, J., Alfs, D., Bednarski, T., Czerwinski, E., et al.: Measurement of gamma quantum interaction point in plastic scintillator with wls strips. Nucl. Instrum. Methods Phys. Res. A 851, 39–42 (2017). doi:10.1016/j.nima.2017.01.045
35. Van Sluis, J., Jong, J., Schaar, J., et al.: Performance characteristics of the digital biograph vision pet/ct system. J Nucl Med. 60(7), 1031–1036 (2019). doi: 10.2967/jnumed.118.215418.
36. Carlier, T., Ferrer, L., Conti, M., et al.: From a pmt-based to a sipm-based pet system: a study tovdefine matched acquisition/reconstruction parameters and nema performance of the biograph vision 450. EJNMMI Phys 7:55 (2020). doi: 10.1186/s40658-020-00323-w.
37. Prenosil, G., Sari, H., Furstner, M., et al.: Performance characteristics of the biograph vision quadra pet/ct system with long axial field of view using the nema nu 2-2018 standard. J Nucl Med 63(3), 476–484 (2021). doi: 10.2967/jnumed.121.261972.
38. Tan, H., Gu, Y., Yu, H., et al.: Total-body pet/ct: Current applications and future perspectives. American Journal of Roentgenology 215(2), 325–337 (2021). doi:10.2214/ajr.19.22705.
39. Petersen, E., Neilson, P., LaBella, A.: Performance simulation of high resolution and high sensitivity prism-pet brain scanner. J Nucl Med 61(1), 1502–1502 (2020)
40. Dadgar, M., Parzych, S., Tayefi, F.: A simulation study to estimate optimum lor angular acceptance for the image reconstruction with the total-body j-pet. In: Smith, Y. (ed.) Medical Image Understanding and Analysis Lecture Notes in Computer Science, Lect. Notes Comput. Sci. ; Oxford, pp. 189–200 (2021). doi:10.1007/978-3-030-80432-915.
41. Kaplon, L.: Technical attenuation length measurement of plastic scintillator strips for the total-body j-pet scanner. IEEE Trans Nucl Sci 67(10), 2286–2289 (2020). doi:10.1109/TNS.2020.3012043.
42. Moskal, P., Stepien, E.: Positronium as a biomarker of hypoxia. Bio-Algorithms and Med-Systems 17(4), 311–319 (2021). doi:10.1515/bams-2021-0189.